\documentclass[prd,superscriptaddress,amsfonts,amssymb,amsmath,showpacs]{revtex4-2}
\usepackage{bm}
\usepackage{amsfonts}
\usepackage{latexsym}
\usepackage[latin1]{inputenc}
\usepackage{graphicx}
\usepackage{amsmath}
\usepackage{palatino}
\usepackage{mathpazo}
\usepackage{textcomp} 
\usepackage{float}
\usepackage{booktabs}
\usepackage{dcolumn}
\usepackage{ragged2e}
\usepackage{hyperref}
\hypersetup{colorlinks,citecolor=green}
\hypersetup{colorlinks=true,linkcolor=blue,filecolor=blue,    urlcolor=magenta}
\usepackage{amsmath}
\usepackage{xcolor}
\usepackage{orcidlink}
\usepackage[caption=false]{subfig}
\usepackage{commath}
\def\jnl@style{\it}
\def\aaref@jnl#1{{\jnl@style#1}}

\def\aaref@jnl#1{{\jnl@style#1}}

\def\aj{\aaref@jnl{AJ}}                   
\def\apj{\aaref@jnl{ApJ}}                 
\def\apjl{\aaref@jnl{ApJ}}                
\def\apjs{\aaref@jnl{ApJS}}               
\def\apss{\aaref@jnl{Ap\&SS}}             
\def\aap{\aaref@jnl{A\&A}}                
\def\aapr{\aaref@jnl{A\&A~Rev.}}          
\def\aaps{\aaref@jnl{A\&AS}}              
\def\mnras{\aaref@jnl{Mon.~Not.~Roy.~Astron.~Soc.}}             
\def\prd{\aaref@jnl{Phys.~Rev.~D}}        
\def\plb{\aaref@jnl{Phys.~Lett.~B}}        
\def\prc{\aaref@jnl{Phys.~Rev.~C}}  
\def\prl{\aaref@jnl{Phys.~Rev.~Lett.}}    
\def\qjras{\aaref@jnl{QJRAS}}             
\def\skytel{\aaref@jnl{S\&T}}             
\def\ssr{\aaref@jnl{Space~Sci.~Rev.}}     
\def\zap{\aaref@jnl{ZAp}}                 
\def\nat{\aaref@jnl{Nature}}              
\def\aplett{\aaref@jnl{Astrophys.~Lett.}} 
\def\apspr{\aaref@jnl{Astrophys.~Space~Phys.~Res.}} 
\def\physrep{\aaref@jnl{Phys.~Rep.}}      
\def\physscr{\aaref@jnl{Phys.~Scr}}       
\def\commat{\aaref@jnl{Comm.~Math.~Phys.}}              
\def\science{\aaref@jnl{Science}}               
\def\cqg{\aaref@jnl{Classical Quant.~Grav.}}            
\def\jpcs{\aaref@jnl{JPCS}}                                     
\def\ijmpd{\aaref@jnl{Int.~J.~Mod.~Phys.~D}}                    
\def\grg{\aaref@jnl{Gen.~Relat.~Gravit.}}               
\def\rpp{\aaref@jnl{Rep.~Prog.~Phys.}}          
\def\npa{\aaref@jnl{Nucl.~Phys.~A}}        
\def\lrr{\aaref@jnl{Living Rev.~Rel.}}                   
\def\jcap{\aaref@jnl{J.~Cosmology Astropart.~Phys.}}    
\def\rmp{\aaref@jnl{Rev.~Mod.~Phys.}}   
\def\epjc{\aaref@jnl{Eur.~Phys.~J.~C}}

\allowdisplaybreaks[1]

\begin{document}

\color{black}       

\title{Traversable wormholes in Rastall-Rainbow Gravity }

\author{Takol Tangphati}  
\email{takoltang@gmail.com}
\affiliation{
School of Science, Walailak University, Thasala, \\Nakhon Si Thammarat, 80160, Thailand}

\author{C. R. Muniz}
\email{celio.muniz@uece.br}
\affiliation{Universidade Estadual do Cear\'a (UECE), Faculdade de Educa\c c\~ao, Ci\^encias e Letras de Iguatu, 63500-000, Iguatu, CE, Brazil.}

\author{Anirudh Pradhan}
\email[]{pradhan.anirudh@gmail.com}
\affiliation{Centre for Cosmology, Astrophysics and Space Science, GLA University, Mathura-281 406, Uttar Pradesh, India}

\author{Ayan Banerjee} 
\email{ayanbanerjeemath@gmail.com}
\affiliation{Astrophysics and Cosmology Research Unit, School of Mathematics, Statistics and Computer Science, University of KwaZulu--Natal, Private Bag X54001, Durban 4000, South Africa}


\date{\today}

\begin{abstract}

In this paper, we investigate the existence of asymptotically flat wormhole geometries within the framework of Rastall-Rainbow modified gravity, a synthesis of two distinct theoretical models: Rastall theory and the Rainbow description. Our study uncovers that, when considering specific combinations of free parameters and equations of state, the emergence of static and spherically symmetric wormholes is not feasible within a zero-tidal-force context. By considering the subset of viable solutions, we conduct a rigorous assessment of their stability through adiabatic sound velocity analysis and scrutinize their compliance with the Weak Energy Condition (WEC). In summary, our inquiry provides insights into how the interplay between Rastall parameters and Rainbow functions may alleviate violations of energy conditions in these modified gravity scenarios.

\end{abstract}

\maketitle

\section{Introduction}

Wormholes are a fascinating theoretical phenomenon predicted by the general theory of relativity (GR). A wormhole is a hypothetical handle-like structure that connects two distant regions in our universe or two different universes.
Historically, the existence of a wormhole solution was first predicted by Austrian physicist Ludwig Flamm, in 1916,  soon after the proposal of GR by Albert Einstein.  Almost twenty years later, Einstein and Rosen \cite{Einstein:1953tkd} suggested the presence of a bridge between asymptotic regions of a two-sided Schwarzschild black hole. This mathematical representation of physical space is technically known as the Einstein-Rosen bridge. In 1988,  Morris and Thorne \cite{Morris:1988cz} proposed an idea that represents a static traversable wormhole connecting two asymptotically flat spacetimes. This study is the cornerstone of a traversable wormhole that can be used for rapid interstellar travel by advanced civilizations.  Subsequently, Morris, Thorne, and Yurtsever \cite{Morris1988} showed the possibility of converting a wormhole into a time machine. All these seminal works put wormholes into an active area of research in GR or in alternative theories of gravitation. Researchers may also see Refs. \cite{Visser:1995,Lobo:2007zb} for general reviews of Lorentzian wormholes in depth.

Another central aspect of a wormhole is the violations of the energy conditions within the GR context at least in a neighborhood of the wormhole throat \cite{Morris:1988cz,Lobo:2007zb}. This means that one needs an amount of exotic matter  (the stress-energy tensor (SET)  of matter violates the null energy condition (NEC) ) to hold the wormhole throat open. As a consequence, the energy density of matter may be seen as negative, at least in some reference frames. It goes without saying that constructing wormholes with the use of the minimum amount of exotic matter has received considerable attention \cite{Visser:2003yf}. In \cite{Visser:2003yf}, authors have theoretically shown that the exotic matter can be made infinitesimally small and confined at the wormhole throat by appropriately choosing the wormhole geometry. The mathematical construction is called cut and paste technique, and the wormhole is called as thin-shell wormhole. The study of thin-shell wormholes has been found in Refs. \cite{Visser:1989kh,Visser:1989kg,Lobo:2003xd,Dias:2010uh}. Further improved quantifier has been proposed by Nandi \textit{et al} \cite {Nandi:2004ku} to know the exact quantity of exotic matter in a given spacetime.

As a result, many arguments have been put forward in support of the violation of the energy conditions. In this direction phantom energy EoS has been used to sustain traversable wormholes \cite{Sushkov:2005kj,Lobo:2005us,Gonzalez:2009cy}.  An interesting point to consider is that in the phantom regime the energy density increases with time and thus provides a notion for the existence of wormholes. Stability analysis of a phantom wormhole geometry has been addressed in \cite{Lobo:2005yv}. Although a variety of wormhole solutions were found for generalized Chaplygin gas \cite{Lobo:2005vc,Kuhfittig:2009mx,Sharif:2014opa}, varying cosmological constant \cite{Rahaman:2006xa}, polytropic phantom energy \cite{Jamil:2010ziq}  and ghost scalar fields \cite{Carvente:2019gkd}. Later, the phantom energy EoS was applied to construct exact evolving wormhole geometries in Ref. \cite{Cataldo:2008pm}. In such a theory, it was found that the phantom energy can support
the existence of evolving wormholes.

However, physicists are always trying to avoid violations of energy conditions or justifying them in a proper way. But to date, it is not possible to construct a static wormhole geometry to satisfy the energy conditions. Thus, researchers are looking for different strategies to alleviate the problem. This observation triggered the possible existence of wormhole solution in the alternative theories of gravity, for example in the higher order gravity theories \cite{Hochberg:1990is,Ghoroku:1992tz}, higher-dimensional cosmological wormholes  \cite{Zangeneh:2014noa} and in the  Einstein-Gauss-Bonnet theory \cite{Bhawal:1992sz,Maeda:2008nz,Mehdizadeh:2015jra}. Considering $f(R)$ gravity, one may theoretically construct traversable wormholes without resorting to exotic matter \cite{Pavlovic:2014gba,Lobo:2009ip}, or sourced by dark matter \cite{Muniz:2022eex}. Otherwise, one can search for wormholes in third-order Lovelock gravity \cite{KordZangeneh:2015dks,Mehdizadeh:2016nna}, hybrid metric-Palatini gravity \cite{Rosa:2021yym,KordZangeneh:2020ixt}, $f(Q)$ gravity \cite{Banerjee:2021mqk,Parsaei:2022wnu,Hassan:2022hcb} and Extended theories of gravity \cite{DeFalco:2021klh,DeFalco:2021ksd}. Traversable wormholes in $f(R,T)$ gravity were studied in Refs. \cite{Moraes:2017mir,Elizalde:2018frj,Moraes:2019pao}. 
At the same time, authors in \cite{Zubair:2019uul,Rosa:2022osy,Banerjee:2020uyi} have found 
exact wormholes solutions without exotic matter in $f(R,T)$ gravity. 

In what follows we shall study the wormhole solutions in recently proposed Rastall-Rainbow gravity  \cite{Mota:2019zln}. The Rastall-Rainbow gravity theory is an alternative theory of gravity which is the combination of Rastall \cite{Rastall:1972swe} and Rainbow theories \cite{Magueijo:2002xx}. Within this context, the structure of NSs was investigated using modern relativistic equations of state (EoS) of nuclear matter derived from relativistic mean field models \cite{Mota:2019zln}. In this work, authors have analyzed separately the effects of each component
on the maximum mass and radius relation of NSs. In \cite{Mota:2022zbq} the authors investigated the anisotropic effects of Rastall and Rainbow parameters 
to describe NSs and QSs. The white dwarfs have particularly been considered in Ref. \cite{Li:2023fux}. A
comprehensive analysis of charged anisotropic strange stars has been studied in Ref. \cite{Das:2022vxq}. In the present paper, we explore the existence of traversable wormhole solutions in the specific theory of Rastall-Rainbow gravity and investigate the effects of the Rastall-Rainbow parameters on the geometric structure. 

Our plan in the present article is the following: In Section \ref{sec2}, we briefly review the newly proposed modified gravity theory, namely, the Rastall-Rainbow gravity. In the same Section, we derive the field equations for traversable wormholes using static and spherically symmetric metrics. In Section \ref{sec3}, several wormhole solutions are
presented, more specifically,  by considering particular choices for EoS relating the energy-momentum tensor components.
Finally,  Section \ref{sec4} is devoted to summarize and discuss our results.


\section{Rastall-Rainbow gravity and stellar structure equations}\label{sec2}

\subsection{Rastall-Rainbow theory}
\label{TOV_RR}

The Rastall-Rainbow gravity model \cite{Mota:2019zln} represents a consistent synthesis of two modified theories of gravity that extend General Relativity (GR): the Rastall theory \cite{Rastall:1972swe} and the Rainbow gravity \cite{Magueijo:2002xx}. The history of the latter begins in 2004 when Jo\~ao Magueijo and Lee Smolin proposed a remarkable generalization of nonlinear special relativity to curved spacetime. By adjusting the formalism that characterizes the principles of this relativity, they successfully incorporated curvature, giving rise to the concept known as 'double general relativity.' This novel approach provided a fresh perspective on spacetime and its dynamics.

An essential consequence of this theoretical work was the discovery of 'Rainbow gravity,' a modification of the standard relativistic dispersion relation $E^{2} - p^{2} = m^{2}$ in the high-energy regime. The modification introduces two arbitrary functions (Rainbow functions) $\Xi(x)$ and $\Sigma(x)$, captured in the equation:
\begin{equation}
E^{2} \Xi(x)^{2} - p^{2}\Sigma(x)^{2} = m^{2}.
\label{eq1}
\end{equation}
In this expression, $x = E/E_{p}$ represents the dimensionless ratio between the energy of the test particle $E$ and a threshold energy usually considered as being the Planck one, $E_{p} = \sqrt{\frac{\hslash c^{5}}{G}}$. This energy is a quantity that is fundamental in physics, determining a natural scale at which quantum gravitational effects become significant. Thus, the functions $\Xi(x)$ and $\Sigma(x)$, with specific functional forms inspired by high-energy phenomena, play a pivotal role in the Rainbow gravity framework, introducing a profound dependence on the test particle energy into the spacetime geometry. Consequently, as particles reach extreme energies near the Planck scale, their motion is substantially influenced by these energy-dependent metrics, leading to relevant phenomena in the behavior of spacetime, emulating backreaction effects. On the other hand, by considering the low energy approximation, $ \Xi(x)$ and $\Sigma(x)$ are chosen so that $x = E/E_{p} \rightarrow 0 $, the usual dispersion ratio is recovered to satisfy the relations: 
\begin{equation}
    \lim_{x\rightarrow 0} \Xi(x)=1, \quad \lim_{x \rightarrow 0} \Sigma(x)=1.
    \label{eq2}
\end{equation}
%

In this context, the spacetime is described using an energy-dependent metric \cite{Magueijo:2002xx} given by:
\begin{equation}
g^{\mu\nu}(x)=\eta^{ab} e_{a}^{\mu}(x)\otimes e_{b}^{\nu}(x),
\label{eq3}
\end{equation}
where $e_{a}^{\mu}(x)$ are the energy-dependent vierbein fields related to the independent ones denoted by $\widetilde{e}_{a}^{\mu}$ through the following expressions:
\begin{equation}
e_{0}^{\mu}(x)=\frac{1}{\Xi(x)} \widetilde{e}_{0}^{\mu}, \quad e_{k}^{\mu}(x)=\frac{1}{\Sigma(x)} \widetilde{e}_{k}^{\mu}.
\label{eq4}
\end{equation}
Here, index $k = (1, 2, 3)$ represents the spatial coordinates. In this way, Einstein's field equations are modified by Rainbow gravity, where the underlying assumption is that the spacetime geometry is dependent on the energy of the test particle (EPT). As a result, all the quantities involved in the field equations in this gravity theory become energy-dependent. In contrast to the traditional Einstein's field equations, a family of other field equations is introduced, with a representative example being:
\begin{equation}
G_{\mu\nu}(x) \equiv R_{\mu\nu}(x) - \frac{1}{2}g_{\mu\nu}(x)R(x) = k(x)T_{\mu\nu}(x),
\label{eq10}
\end{equation}
where $k(x) = 8 \pi G(x)$.
Such modifications are of significant interest and lead to novel insights into the interplay between gravity, high-energy physics, and the underlying structure of spacetime. 

Peter Rastall made a significant contribution to the gravitational theory in the early 1970s by proposing a modification to the conservation principles of the energy-momentum tensor in curved spacetime \cite{Rastall:1972swe}. His groundbreaking work aimed to retain the validity of the gravitational Bianchi identity of the Einstein geometric tensor while incorporating necessary adjustments. Rastall's idea was to consider that the divergence of the energy-momentum tensor ($T_{\mu\nu}$) was proportional to the variation of the Ricci scalar ($R$), suggesting that geometry also must contribute to the total energy of the system. Some arguments have been raised to claim that Rastall gravity is entirely equivalent to standard Einstein gravity \cite{Visser:2017gpz}, Nevertheless, the validity of this proposition has been called into question by several studies \cite{Darabi:2017coc}. 

The proposed modified conservation law by Rastall is expressed as \cite{Mota:2019zln}: 
\begin{equation}
\nabla^{\mu}T_{\mu\nu} = \bar{\lambda}\nabla_{\nu}R,
\label{eq6}
\end{equation}
where $\bar{\lambda} = \frac{1 - \lambda}{16 \pi G}$, and $\lambda$ is the Rastall parameter representing the coupling between geometry and matter fields \cite{Das:2018dzp}. When $\bar{\lambda} = 0$ (or $\lambda = 1$),  GR is recovered. In flat spacetime, where $R$ vanishes, the conventional conservation law for $T_{\mu\nu}$ is restored. Thus, the modified Einstein field equations consistent with equation (\ref{eq6}) can be written as \cite{Mota:2019zln}: 
\begin{equation}
R_{\mu}{}^{\nu} - \frac{\lambda}{2}\delta_{\mu}{}^{\nu}R = 8\pi G T_{\mu}{}^{\nu},
\label{eq8}
\end{equation}

The Rastall-Rainbow model unifies the effects of Rainbow gravity with the effects of Rastall gravity, under a single formalism. The field equations in this unified formalism are written as follows:
\begin{equation}
R_{\mu}{}^{\nu}(x) - \frac{\lambda}{2}\delta_{\mu}{}^{\nu}(x)R(x) = k(x) T_{\mu}{}^{\nu}(x)  
 \label{eq7},
\end{equation}
where the Rastall parameter $\lambda$ is independent of the test particle energy. In the following discussion, our object is to represent the simplest wormhole geometry and discuss the features of its geometry.

\subsection{The wormhole geometry and the field equations}

In order to study the wormhole solution, we begin with a simple spherically symmetric and time-independent (static) metric. In such manner, we replace the usual GR quantities $\widetilde{e}_{i}$ for spherical symmetry into Eq. (\ref{eq3}), which yield
\begin{equation}
    ds^{2}=-\frac{e^{2\Phi(r)}}{\Xi^{2}(x)} dt^{2}+ \frac{dr^2}{\Sigma^{2}(x) \left(1-\frac{b(r)}{r}\right)}+\frac{r^{2}}{\Sigma^{2}(x)}(d\theta^{2}+\sin{\theta}^{2}d\phi^{2}),
    \label{eq5}
\end{equation}
where $\Phi(r)$ and $b(r)$ are arbitrary functions of the radial coordinate and are known as the redshift and the shape functions, respectively.  We can see that the metric potentials are depending on the rainbow functions $\Xi(x)$ and $\Sigma(x)$. However, the coordinates $r$, $t$, $\theta$, and $\phi$ are independent of the energy probe particles in gravity's rainbow. A general property of wormhole physics is the existence of a throat connecting two asymptotically flat spacetime with
a minimum value surface radius $b(r_0)= r_0$. Since, the radial coordinate covering the range of two patches $[r_0, +\infty)$, 
and $r_0$ is the passage between two patches. On the other hand, flaring-out condition is another fundamental ingredient
of wormhole geometry, and the condition is given by  $\frac{b(r)-rb^{\prime}(r)}{b^2(r)}>0$ \cite{Morris:1988cz}. 
More precisely, the form function satisfies the condition $b^{\prime}(r_0) < 1$. In addition to this one needs to impose the condition $1-b(r)/r > 0$ for the region out of the throat. The redshift function $\Phi(r)$ is to be finite throughout the spacetime to avoid the presence of event horizons.  

Our aim is to model a wormhole solution and typically we consider an anisotropic fluid stress-energy tensor to describe the matter distribution, which is described by the following form
\begin{equation}\label{eq12}
T_{\mu\nu}=(\rho+p_t)u_\mu u_\nu+ p_t g_{\mu\nu}-(p_{t}-p_{r}) \chi_{\mu}\chi_{\nu},
\end{equation}
where $u^\mu$ is the fluid 4-velocity ($u_\mu u^\mu = -1$), $\chi_{\mu}$ is the unit radial vector so that $\chi_{\mu} \chi^{\mu} = 1$ and $g_{\mu\nu}$ is the metric tensor. Furthermore, $\rho = \rho(r)$ is the energy density,  $p_r = p_r(r)$ and $p_t = p_t (r)$ are the radial and tangential pressures, respectively.

Now, rewriting the Eq. (\ref{eq8}) in its covariant form we construct the geometrical Einstein tensor on the left-hand side with an effective energy-momentum tensor on the right-hand side, as
\begin{equation}
     R_{\mu\nu}-\frac{1}{2}g_{\mu\nu}R=8\pi G\tau_{\mu \nu},
     \label{eq13b}
 \end{equation}
  where
 \begin{equation}
     \tau_{\mu\nu}=T_{\mu\nu}-\frac{(1-\lambda)}{2(1-\lambda)}g_{\mu\nu}T.
      \label{eq13c}
 \end{equation}
 Taking into account the Eqs. (\ref{eq5}) and (\ref{eq12}), the nonzero  components of the field equations (\ref{eq13b}) reduce to
  \begin{eqnarray}
&& \frac{b^{\prime}}{r^{2}} = 8\pi  \Bar{\rho},  \label{eq14} \\
&& 2\left(1-\frac{b}{r}\right)\frac{\Phi^{\prime}}{r}-\frac{b}{r^{3}}  = 8\pi  \Bar{p}_{r},  \label{eq15} \\
&& \left(1-\frac{b}{r}\right)\left[\Phi^{\prime\prime}+\Phi^{\prime 2}-\frac{b^{\prime}r-b}{2r(r-b)}\Phi^{\prime}-\frac{b^{\prime}r-b}{2r^2(r-b)} +\frac{\Phi^{\prime}}{r}\right] 
 = 8\pi  \Bar{p}_{t},  \label{eq16}
\end{eqnarray}
where $\Bar{\rho}$ represents the effective energy density, $\Bar{p}_{r}$ is the effective radial pressure and $\Bar{p}_{t}$ is the effective tangential pressure respectively, we define
\begin{align}
    \Bar{\rho} & = \frac{1}{\Sigma(x)^{2}}\left[\alpha_{1}\rho+\alpha_{2}p_{r}+2\alpha_{2}p_{t}\right],\label{eq17}\\
    \Bar{p}_{r} & = \frac{1}{\Sigma(x)^{2}}\left[\alpha_{2}\rho+\alpha_{1}p_{r}-2\alpha_{2}p_{t}\right], \label{eq18}\\
    \Bar{p}_{t} & = \frac{1}{\Sigma(x)^{2}}\left[\alpha_{2}\rho-\alpha_{2}p_{r}+\alpha_{3}p_{t}\right], \label{eq18b}
\end{align}
with
\begin{equation*}
    \alpha_{1}=\frac{1-3\lambda}{2(1-2\lambda)}, \qquad \alpha_{2}=\frac{1-\lambda}{2(1-2\lambda)}, \qquad
    \alpha_{3}=-\frac{\lambda}{1-2\lambda}.
\end{equation*}
If we fix our choice for $\lambda=1$ and $\Sigma=1$ then one may recover the usual definition of the GR for an anisotropic fluid sphere. The modified field equations (\ref{eq14})$-$(\ref{eq16}), have five unknown quantities, i.e., $\Phi(r)$, $b(r)$, $\rho(r)$, $p_r(r)$ and $p_t(r)$ with three independent differential equations.
Obviously, this is an undetermined system of equations, and for a complete description of the wormhole configuration, one may apply or adopt different strategies. For our purpose, we restrict ourselves
to consider $\Phi(r)$ and $b(r)$, respectively. Finally, we can write the conservation equation of the energy-momentum tensor $T^{\nu}_{\:\:\:\mu;\nu}=\Bar{\lambda}R_{\mu}$ takes the following form
\begin{equation}
 \Bar{p}_{r}' = -(\Bar{p}_{r}+\Bar{\rho})\Phi'+ \frac{2}{r}\left(\Bar{p}_{t}-\Bar{p}_{r}\right).
 \label{eq21}
\end{equation}


\section{Specific Wormhole Model} \label{sec3}

In the following analysis, as explained hereunder, we consider the zero-tidal wormholes, {\it i.e.} those ones that present a constant redshift function i.e., $\Phi'=0$, which simplifies our entire calculations and reduces the number of unknown quantities. 

\subsection{Specific solution: $p_r=\omega \rho$}

\begin{figure}[h]
    \centering
    \includegraphics[width = 7.8 cm]{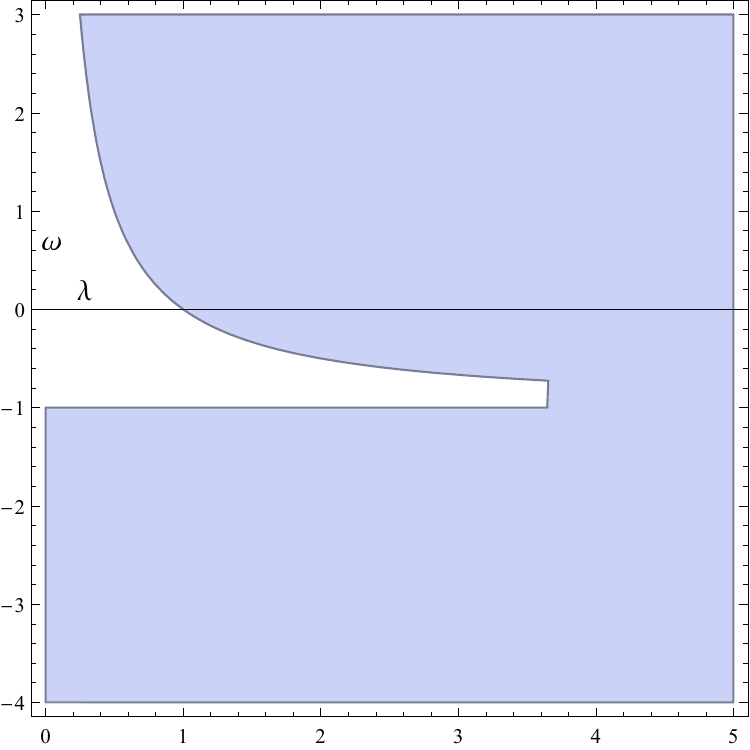}
    \includegraphics[width = 8 cm,height=8.5cm]{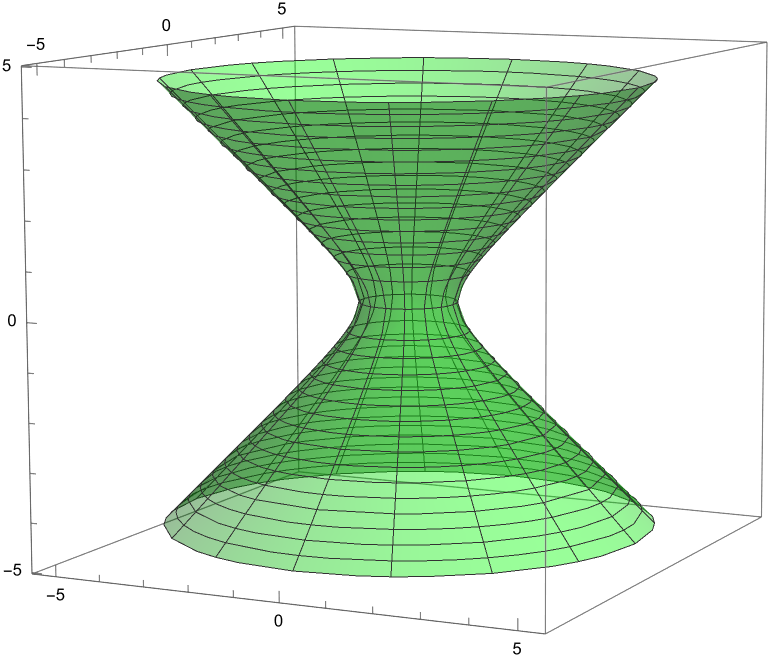}
    \caption{Left Panel: Parameter space ($\lambda,\omega$) indicating regions where the flaring-out conditions are satisfied for the zero-tidal wormhole solution, in blue. Right Panel: The embedding diagram for a wormhole using the Eq. (\ref{embed1}) for $\lambda$ = 0.6 and $\omega$ = 2.0.}
    \label{fig:C01}
\end{figure}

We consider now the state equation $p_r=\omega \rho$ and solve Eqs. (\ref{eq14})-(\ref{eq16}) for $b(r)$, $p_r$, and $p_t$, finding, firstly:
\begin{equation}
    b(r)=r_0\left(\frac{r_0}{r}\right)^{\frac{1}{\lambda  \omega +\lambda -1}}, \label{br_modA}
\end{equation}
where the integration constant was chosen so that $b(r_0)=r_0$. The zero tidal solution for the wormhole remains unaffected by the rainbow function $\Sigma(x)$, implying its robustness regardless of specific forms. To satisfy the condition $1 - \frac{b(r)}{r} > 0$ for $r > r_0$, the constraint $\frac{1}{\lambda \omega + \lambda - 1} + 1 > 0$ must be imposed, and it is crucial to maintain $b'(r_0) < 1$ and $b(r) - b'(r) r >0$, as we already have seen. Fig. \ref{fig:C01} (left panel) illustrates the parameter space ($\lambda, \omega$) and identifies regions where these flaring-out conditions are met, providing valuable insights into the wormhole's existence. We can discern that as $\lambda$ approaches zero, the existence of the wormhole is limited to $\omega < -1$, which corresponds to a phantom-like source. Conversely, for larger values of $\lambda$, all forms of matter can potentially create a wormhole. Notably, these results hold true for any value of the throat radius.  To get a better visualization we take the help of geometrical embedding diagrams.  Following Ref. \cite{Morris:1988cz},  one can directly write the equation for the embedding surface, which is
 \begin{eqnarray}\label{embed1}
\frac{dz}{dr} = \pm \left(\frac{r}{b(r)}-1\right)^{-1/2}.
\end{eqnarray}
By utilizing the Eq. (\ref{embed1}), we plot the embedding diagram through a $2\pi$ rotation around the $z$-axis for $\lambda$ = 0.6 and $\omega$ = 2.0 in Fig. \ref{fig:C01} (right panel). This is also verified that the spacetime is asymptotically flat, i.e., $\frac{dz}{dr}\to 0$ as $r\to \infty$.
\begin{figure}[h]
    \centering
    \includegraphics[width = 8 cm]{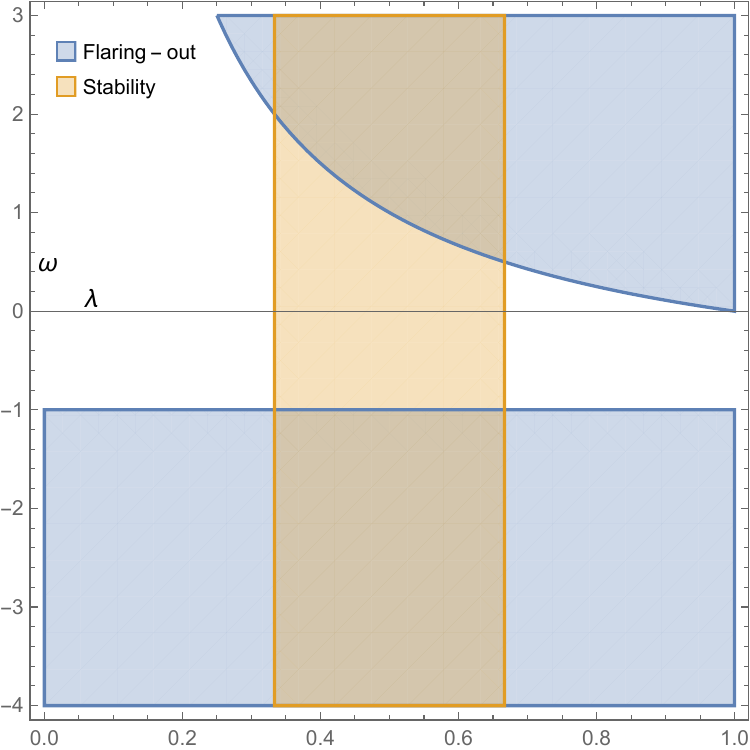}
    \includegraphics[width = 8.2 cm,height=8.3cm]{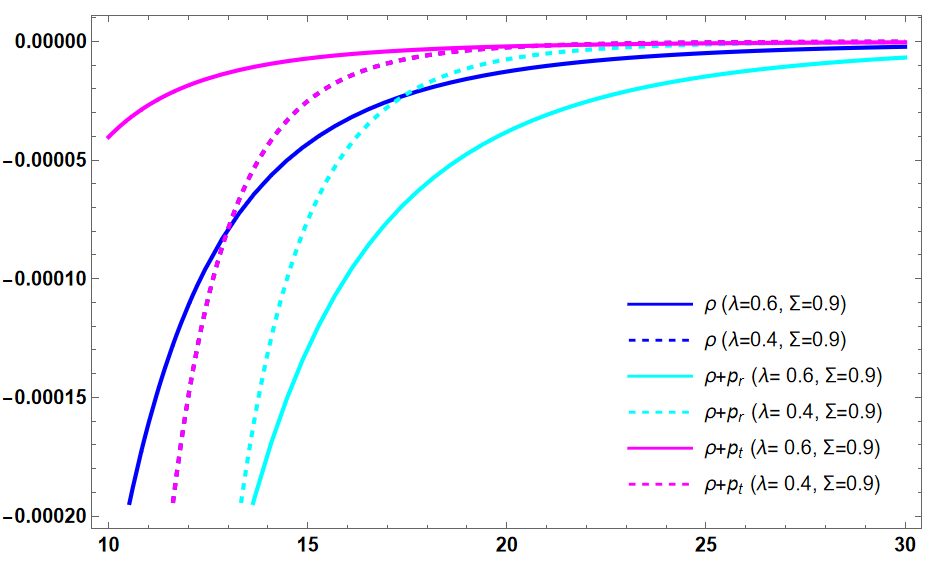}
    \caption{Left panel: Parameter space ($\lambda,\omega$) indicating regions where the flaring-out conditions are satisfied for the zero-tidal wormhole solution, case A, in the blue area and the region satisfying the stability conditions in the orange area. Right panel: Energy density and its combination with the pressures, for some values of
    $r_0=10$, $\omega=2$ and $\Sigma=0.9$, respectively.}
    \label{fig:C02}
\end{figure}
\\

Since we have determined the shape function $b(r)$, the energy density and pressures can easily be found:
\begin{eqnarray}
    \rho(r)&=&-\frac{\lambda  r_0 \Sigma^2 }{8 \pi  r^3 (\lambda  \omega +\lambda -1)}\left(\frac{r_0}{r}\right)^{\frac{1}{\lambda  \omega +\lambda -1}},\label{rho}\\
    p_r(r)&=&\omega \rho(r),\label{p_r}\\
    p_t(r)&=&\frac{r_0 \Sigma^2 (\lambda  \omega +3 \lambda -2) }{16 \pi  r^3 (\lambda  \omega +\lambda -1)}\left(\frac{r_0}{r}\right)^{\frac{1}{\lambda  \omega +\lambda -1}}\label{p_t}.
\end{eqnarray}
The connection between energy density, pressures, and the Rainbow function $\Sigma(x)$ becomes evident. Our capacity to assess energy conditions is now operational. In the right panel of Fig. \ref{fig:C02}, the depicted data clearly illustrates a consistent departure from compliance with both the Weak Energy Condition (WEC) and the Null Energy Condition (NEC) across all energy density and pressure combinations examined. Notably, a reduction in the Rastall parameter, $\lambda$, exacerbates these violations near the wormhole throat while concurrently diminishing them as we move away from the throat. The attenuation is further influenced by the Rainbow function $\Sigma$ as it diminishes in value, due to the direct proportionality of the energy density and pressures with its square. 

Both the stability and physical meaning of the solutions can be effectively assessed using the adiabatic sound velocity, denoted as $v_s^2= \frac{\partial{<p>}}{\partial{\rho}} $, where $<p>$ represents the average pressure across the three spatial dimensions, namely $<p> = \frac{1}{3}(p_r + 2p_t)$. These essential properties remain valid under the constraint:
\begin{equation}
0 \leq v_s^2 < 1.
\end{equation}
By examining the equations (\ref{rho}), (\ref{p_r}), and (\ref{p_t}), we deduce:
\begin{equation}\label{Sound}
v_s^2 = \frac{d<p>}{dr} \left( \frac{d\rho}{dr}\right)^{-1} = \frac{2}{3 \lambda} - 1.
\end{equation}

Consequently, it follows that the parameter $\lambda$ must reside within the interval $\left(\frac{1}{3}, \frac{2}{3}\right)$ to satisfy these conditions. Fig. \ref{fig:C02} (left panel) depicts the parameter space highlighting the regions in which both the flaring-out and stability conditions are simultaneously obeyed.


\subsection{Specific solution: $p_t= m \rho$ }

\begin{figure}[h]
    \centering
    \includegraphics[width = 8 cm]{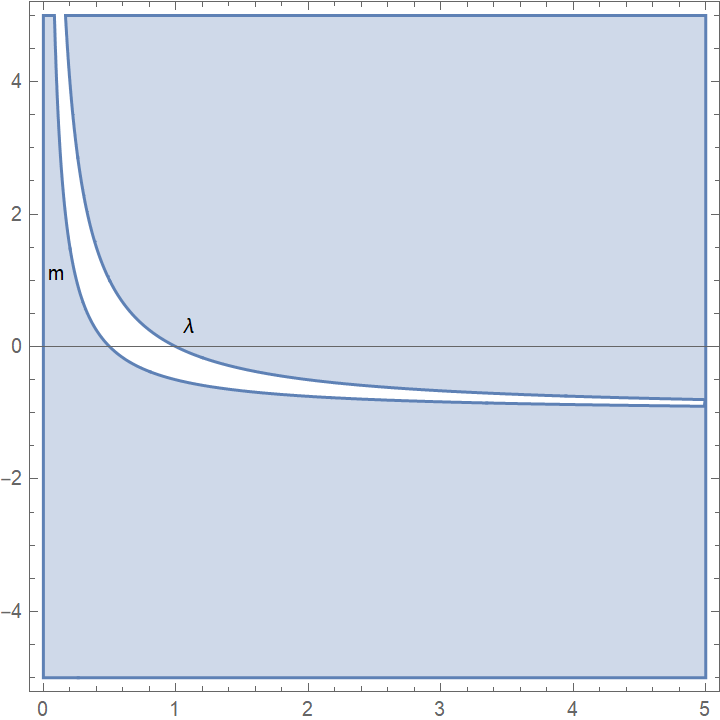}
    \includegraphics[width = 8 cm,height=8.5cm]{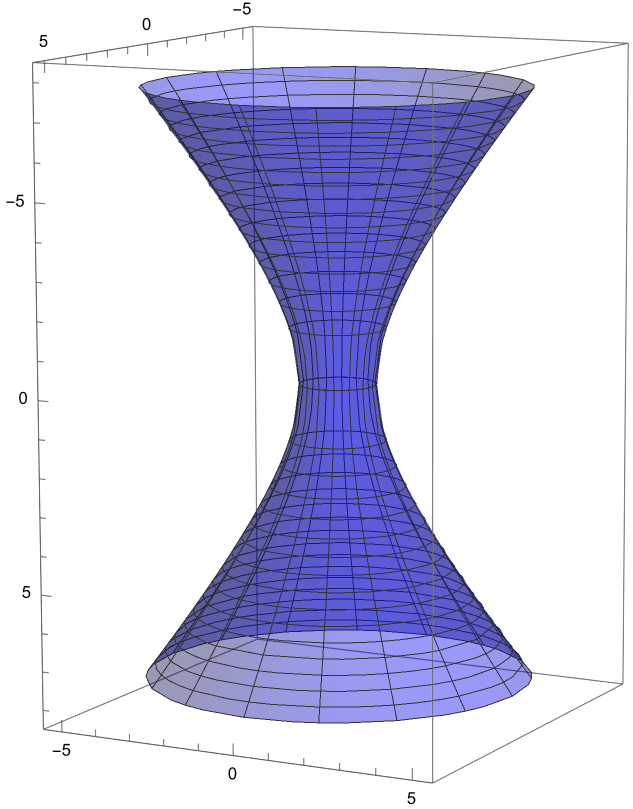}
    \caption{Left Panel: Parameter space ($\lambda, m$) indicating regions where the flaring-out conditions are satisfied for $\Phi'(r)=0$ and $p_t= m \rho$, in blue. 
    Right Panel: The embedding diagram for a wormhole using Eq. (\ref{embed1}) and  Eq. (\ref{br_modB}) for $\lambda$ = 0.6 and $m$ = 2.0.}
    \label{fig:B01}
\end{figure}

Here, we consider another specific form of EoS $p_t= m \rho$ where  $m$ is the state parameter. For this case the  shape function is given by
\begin{equation}
b(r)=r_0 \left(\frac{r_0}{r}\right)^{\frac{1}{1-2 \lambda  (m+1)}}. \label{br_modB}
\end{equation}
where we recall $b(r_0)=r_0$ to fix the integration constant. In this case, also the obtained shape function is  unaffected by the rainbow function $\Sigma(x)$ and depends only on the Rastall parameter
$\lambda$. As of the previous case we impose the conditions $1+\frac{1}{1-2 \lambda  (m+1)}>0 $ and $(\lambda +\lambda  m-1) (2 \lambda  (m+1)-1)>0$  to satisfy the  asymptotic flatness and flaring-out conditions.  Fig. \ref{fig:B01} (left panel) shows the parameter space ($\lambda, m$) where the flaring-out conditions are met in support of the wormhole's existence. Observing the parameter space,  we come to know that one may theoretically construct traversable wormholes in most of the region especially when $m<0$. Interestingly, 
these results are true for any value of the throat radius. Now, using the Eq.  (\ref{embed1}) and Eq. (\ref{br_modB}), one finds the embedding surface as is depicted in Fig. \ref{fig:B01} (right panel), which confirms the 
asymptotically flat spacetime, i.e., $\frac{dz}{dr}\to 0$ as $r\to \infty$ for $\lambda$ = 0.6 and $m$ = 2.0.

With the Eq.  (\ref{br_modB}), the nonvanishing components of $ T_{\mu \nu}$ are
\begin{eqnarray}
\rho (r)&=& \frac{\lambda  r_0 \Sigma^2 }{8 \pi  r^3 (2 \lambda  (m+1)-1)}\left(\frac{r_0}{r}\right)^{\frac{1}{1-2 \lambda  (m+1)}}, \label{rho2} \\
p_r(r)&=& -\frac{r_0 (\lambda  (2 m+3)-2) \Sigma^2 }{8 \pi  r^3 (2 \lambda  (m+1)-1)}\left(\frac{r_0}{r}\right)^{\frac{1}{1-2 \lambda  (m+1)}}, \label{pr2}\\
p_t(r)&=& m \rho. \label{pt2}
\end{eqnarray}

\begin{figure}[h]
    \centering
    \includegraphics[width = 8 cm]{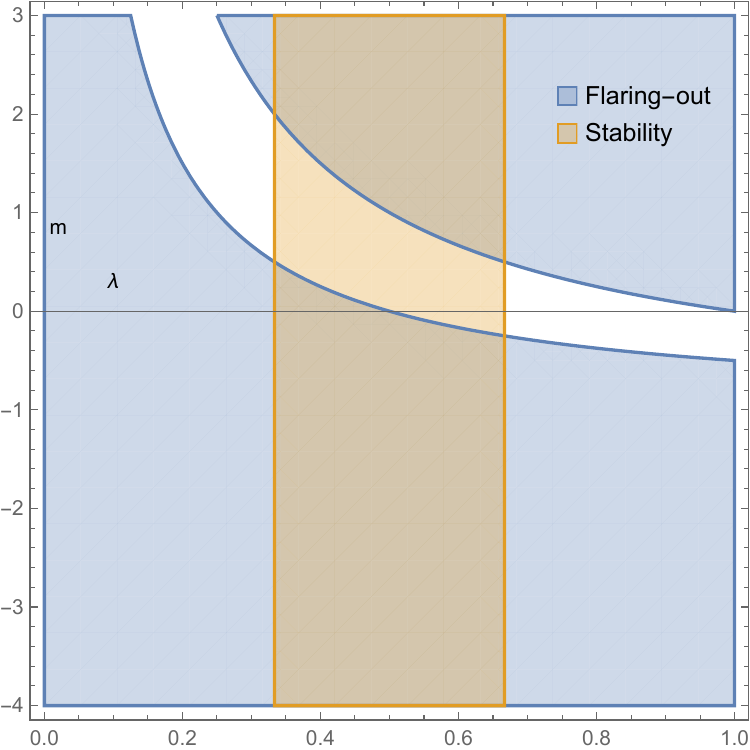}
        \includegraphics[width = 8.4 cm,height=8.3cm]{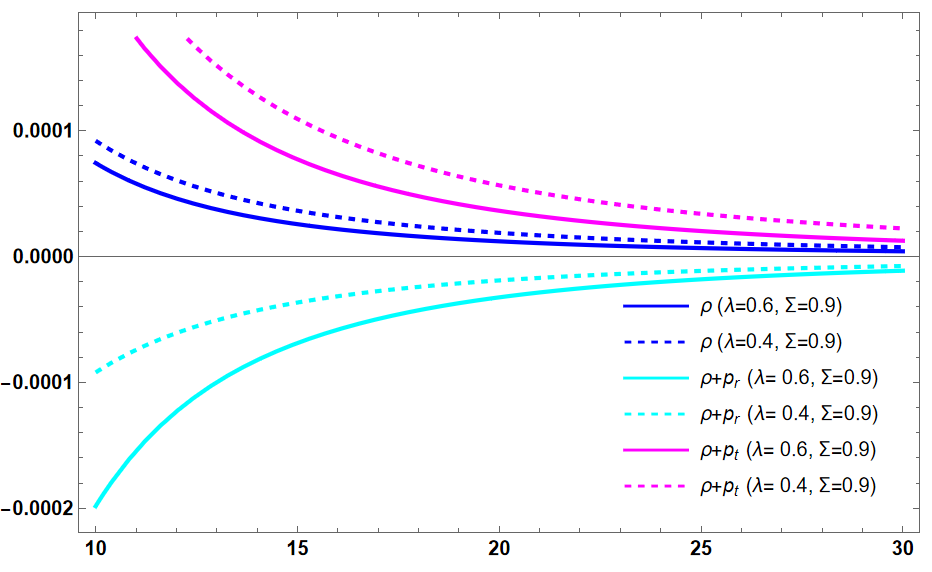}
    \caption{Left panel: Parameter space ($\lambda, m$) for the wormhole solution in which $p_t=m \rho$, indicating regions where the flaring-out conditions, in blue, and stability conditions, in orange, are satisfied. Right Panel: Energy density and its sum with the pressures, as a function of the radial coordinate, for case B and considering some values of $\lambda$, with $r_0=10$, $m=2$, and $\Sigma=0.9$, respectively.}
    \label{fig:B02}
\end{figure}

By examining the equations (\ref{rho2}), (\ref{pr2}), and (\ref{pt2}), we deduce:
\begin{equation}\label{Sound3}
v_s^2 = \frac{d<p>}{dr} \left( \frac{d\rho}{dr}\right)^{-1} = \frac{2}{3 \lambda }-1.
\end{equation}

It is essential to note that the adiabatic sound velocity coincides with the value computed in the previous subsection. This congruence serves as an indicator of stability and physical meaning within the interval of $1/3 \leq \lambda < 2/3$. In Fig. \ref{fig:B02} (left panel), we show the regions where the flaring-out and stability conditions hold simultaneously.

In order to analyze the weak energy conditions - WEC ($\rho+p_i\geq 0$, $\rho\geq$ 0), by choosing parameters that allow the wormhole formation we depict in Fig. \ref{fig:B02} (right panel) some quantities, namely, energy density as well as its sum with the radial and lateral pressures separately, as functions of the radial coordinate, for two values of $\lambda$. We can notice that the reduction of this parameter improves those conditions, principally where they are violated. The same can be said with respect to the rainbow function, given that such quantities are proportional to $\Sigma^2$. In simpler terms, as we distance ourselves from GR by decreasing Rastall's parameters and adjusting the Rainbow functions -- achieved by increasing the energy of the test particle -- the violation of energy conditions is progressively alleviated.

\begin{figure}[h]
    \centering
    \includegraphics[width = 8 cm]{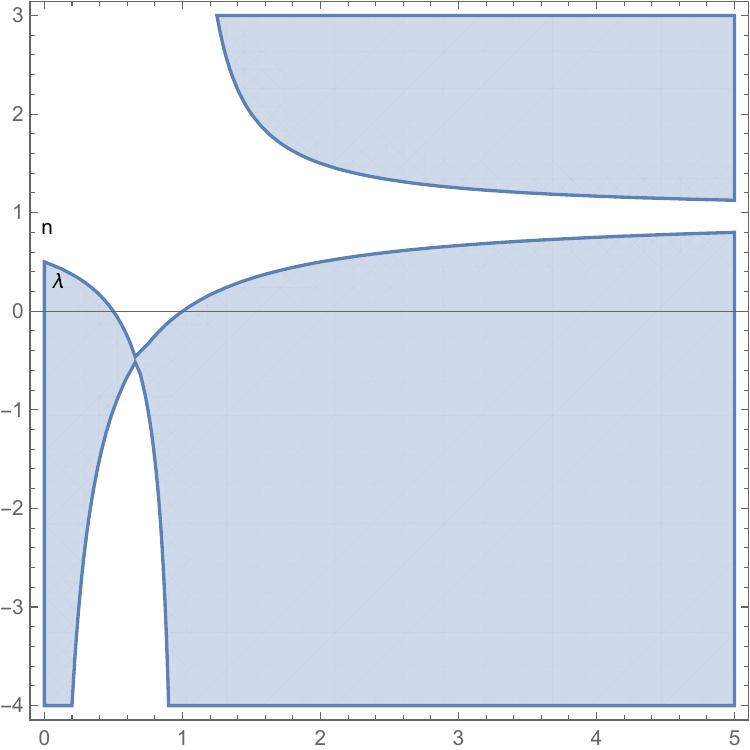}
    \includegraphics[width = 8.3 cm]{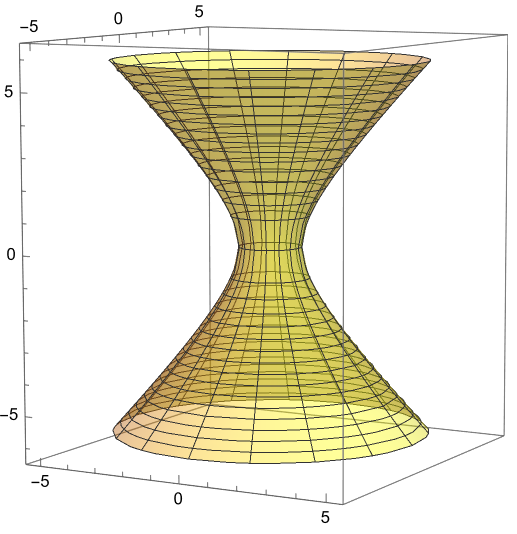}
    \caption{Left Panel: Parameter space ($\lambda,n$) indicating regions where the flaring-out conditions are satisfied for the zero-tidal wormhole solution, in blue. Right Panel: The embedding diagram for a wormhole using the Eq. (\ref{embed1}) and  Eq. (\ref{br_modB}) for 
    $\lambda = 0.4$ and $n = -0.75$.}
    \label{fig:C03}
\end{figure}

\begin{figure}[h]
    \centering
    \includegraphics[width = 8 cm]{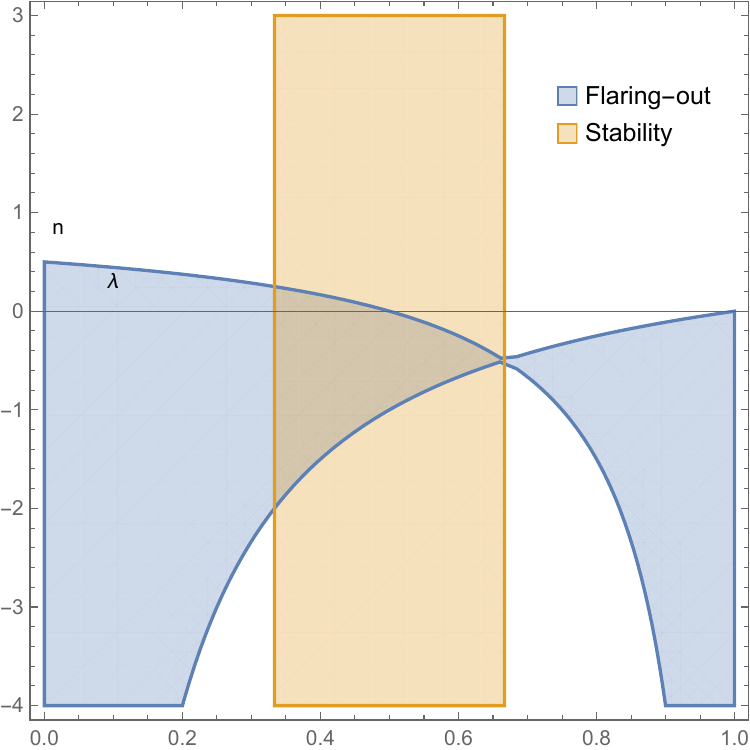}
    \includegraphics[width = 8.3 cm,height=8.5cm]{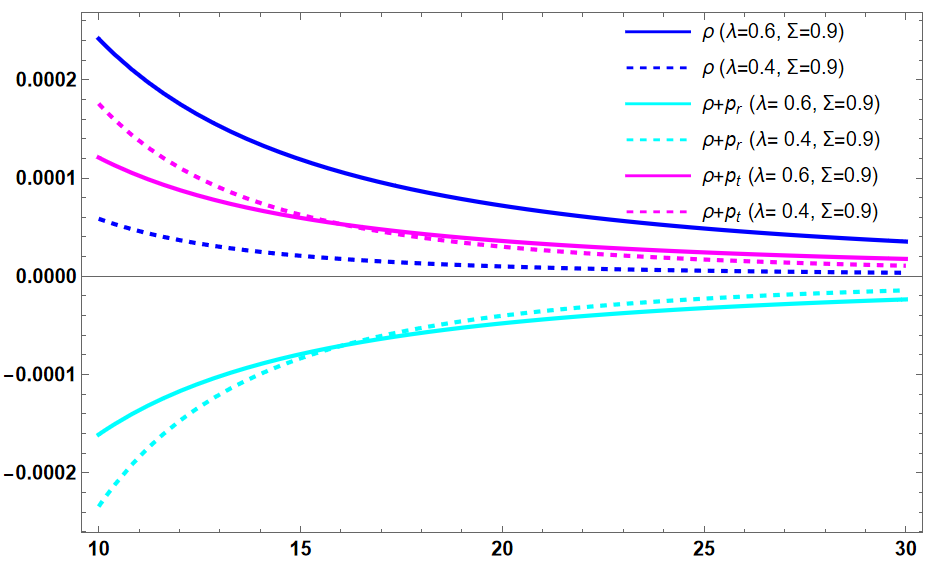}
    \caption{Left Panel: Parameter space ($\lambda, n$) for the wormhole solution in which $p_t=n p_r$, indicating regions where the flaring-out conditions, in blue, and stability conditions, in orange, are satisfied. Right Panel: Energy density and its sum with the pressures, as a function of the radial coordinate, for case C and considering some values of $\lambda$, with $r_0=10$, $n= -0.75$ and $\Sigma=0.9$, respectively.}
    \label{fig:C002}
\end{figure}
\subsection{Specific solution: $p_t = n p_r$ }

To this end, we consider an equation establishing a relation between two pressure components as $p_t = n p_r$ \cite{Rahaman:2006xa,Moraes:2017mir} and solve Eqs. (\ref{eq14})-(\ref{eq16}) for $b(r)$, $p_r$, and $p_t$, which yield
\begin{equation}\label{shpe3}
b(r)=r_0 \left(\frac{r_0}{r}\right)^{\frac{2 n+1}{-2 \lambda +2 (\lambda -1) n+1}},
\end{equation}
using the condition $b(r_0)=r_0$ as an integration constant. Similar to the previous cases, we observe that the obtained shape function (\ref{shpe3}) is free of $\Sigma(x)$. Now, depending on the Eq. (\ref{shpe3}) and applying all the necessary constraints on $b(r)$, we impose two conditions on the free parameter $n$ as $1+\frac{2 n+1}{-2 \lambda +2 (\lambda -1) n+1}>0$ and  $(-2 \lambda +2 (\lambda -1) n+1) (\lambda  (n-1)+1)>0$.
In Fig. \ref{fig:C03} (left panel), we illustrate the parameter space ($\lambda, n$) where the flaring-out condition is met. For this solution, one verifies that when $\lambda$ approaches zero, the existence of wormholes is restricted. On the other hand,  for larger values of $\lambda$, the wormholes can exists mainly for $n< 0$.  In Fig. \ref{fig:C03} (right panel), we depict the embedding surface using Eq.  (\ref{embed1}) and Eq. (\ref{shpe3}) for $\lambda$ = 0.6 and $n$ = - 0.75, respectively.

Using Eqs. (\ref{eq14})-(\ref{eq16}) and the Eq. (\ref{shpe3}), we have the follow stress-energy tensor profile 
\begin{eqnarray}
    \rho(r)&=&-\frac{\lambda  (2 n+1) r_0 \Sigma^2 }{8 \pi  r^3 (-2 \lambda +2 (\lambda -1) n+1)}\left(\frac{r_0}{r}\right)^{\frac{2 n+1}{-2 \lambda +2 (\lambda -1) n+1}} \label{rho3},\\
    p_r(r)&=& \frac{(3 \lambda -2) r_0 \Sigma^2 }{8 \pi  r^3 (-2 \lambda +2 (\lambda -1) n+1)}\left(\frac{r_0}{r}\right)^{\frac{2 n+1}{-2 \lambda +2 (\lambda -1) n+1}}, \label{p_r3}\\
    p_t(r)&=& n p_r. \label{p_t3}
\end{eqnarray}

As we are considering the adiabatic sound velocity, using the relation in (\ref{rho3})-(\ref{p_t3}), we have
\begin{equation}\label{Sound3}
v_s^2 = \frac{d<p>}{dr} \left( \frac{d\rho}{dr}\right)^{-1} = \frac{2}{3 \lambda }-1.
\end{equation} 

Interestingly, the adiabatic sound velocity derived in this context is identical to the results obtained in the previously analyzed cases. Thus, the stability and physicality criteria remain consistent across these scenarios as well. Fig. \ref{fig:C002} (left panel) illustrates the parameter space ($\lambda, n$), highlighting the regions where both the flaring-out and stability conditions are simultaneously satisfied.

As discussed in previous sections,  we now turn to the question of whether it entails WEC violations. The graphical behavior of the WEC is presented in Fig. \ref{fig:C002} (right panel) for two different values of $\lambda$. As we can see from the figure, by decreasing the value of $\lambda$ one can improve the violation of energy conditions. Also, the situation is the same for decreasing values of $\Sigma^2$. Finally, we conclude that the Rastall-Rainbow gravity is expected to diminish the violation of energy conditions depending on the two model parameters.   


\section{Concluding remarks }\label{sec4}
For more than one hundred years, Einstein's theory of general relativity is still a great successful gravity theory 
on solar system scales. However, the missing matter problem at all astrophysical scales and the current accelerated expansion of the universe have motivated us to consider the possibility of modified gravity theories. Numerous modified gravity theories have emerged, each offering unique perspectives on gravitational dynamics while remaining consistent with current observational constraints. Among these alternatives, the relatively recent Rastall-Rainbow gravity theory has garnered attention. This theory arises from the fusion of two distinct frameworks: the Rastall theory, which relaxes the conservation of energy-momentum, and the Rainbow theory, which accounts for the metric's dependence on the energy of probe particles. In this paper, our aim is to introduce and explore the existence of asymptotically flat wormhole geometries that find support within the theoretical framework of Rastall-Rainbow gravity.

In this study, we have derived the gravitational field equations governing static wormhole solutions. We have applied stringent constraints to our model parameters using the flaring-out conditions. Specifically, we have constructed asymptotically flat wormholes by utilizing three equations of state that establish a relationship between density and pressure components, with fixed state parameters. Notably, the resulting shape functions are entirely independent of the rainbow function, relying solely on the Rastall parameter and on the state parameters.

Our investigation further extends to the analysis of the adiabatic sound velocity, a critical factor in determining the overall stability of the fluid structure within the wormhole. Significantly, the sound speed is exclusively dependent on the Rastall parameter and consistently falls within a predetermined range in all examined cases. Parameter spaces were furnished in order to show the conditions of validity for the wormhole formation and its stability.

Finally, we have assessed the Weak Energy Condition (WEC) and found that a careful parameter selection can mitigate violations of the energy conditions. Specifically, the Rainbow functions (independent of its form) always contribute to such mitigation in all space. In other words, a probe particle with very high energy indefinitely reduces the exoticity of the wormhole source on passing near this object. This study provides a comprehensive and objective examination of static wormhole solutions, offering insights into their stability and physical viability within our extended gravity model's framework.
 
\section*{date availability}

There are no new data associated with this article.

\begin{acknowledgments}
 Takol Tangphati is financially supported by Research and Innovation Institute of Excellence, Walailak University, Thailand under a contract No. WU66267.
\end{acknowledgments}\

\end{document}